\documentclass[aps,superscriptaddress,twocolumn,twoside,floatfix,prx,nofootinbib,a4paper]{revtex4-1}
\usepackage{graphicx,amsmath,amssymb,color}
\usepackage[normalem]{ulem}
\usepackage{amsfonts}
\usepackage[ocgcolorlinks,colorlinks=true,linkcolor=blue,citecolor=red]{hyperref}
\usepackage{latexsym}
\usepackage{amsfonts}
\usepackage{mathrsfs}
\usepackage{natbib}
\usepackage{color,verbatim}
\DeclareMathAlphabet{\mathrsfs}{U}{rsfs}{m}{n}
\DeclareMathAlphabet{\mathpzc}{OT1}{pzc}{m}{it}
\DeclareMathAlphabet{\matheus}{U}{eus}{m}{n}
\DeclareMathAlphabet{\mathbbold}{U}{bbold}{m}{n}

\newcommand{\ba}{\begin{eqnarray}}
\newcommand{\be}{\begin{equation}}
\newcommand{\ee}{\end{equation}}

\newcommand{\ea}{\end{eqnarray}}
\newcommand{\ban}{\begin{eqnarray*}}
\newcommand{\ean}{\end{eqnarray*}}

\newcommand{\ket}[1]{|#1\rangle}

\begin{document}

\title{Quantum Communication Complexity using the Quantum Zeno Effect}

\author{Armin Tavakoli}
\affiliation{Department of Physics, Stockholm University, S-10691 Stockholm, Sweden}

\author{Hammad Anwer}
\affiliation{Department of Physics, Stockholm University, S-10691 Stockholm, Sweden}

\author{Alley Hameedi}
\affiliation{Department of Physics, Stockholm University, S-10691 Stockholm, Sweden}
\author{Mohamed Bourennane}
\affiliation{Department of Physics, Stockholm University, S-10691 Stockholm, Sweden}

\begin{abstract}
The quantum Zeno effect (QZE) is the phenomenon where the unitary evolution of a quantum state is suppressed e.g. due to frequent measurements. Here, we investigate the use of the QZE in a class of communication complexity problems (CCPs). Quantum entanglement is known to solve certain CCPs beyond classical constraints. However, recent developments have yielded CCPs where super-classical results can be obtained using only communication of a single $d$-level quantum state (qudit) as a resource. In the class of CCPs considered here, we show quantum reduction of complexity in three ways: using i) entanglement and the QZE, ii) single qudit and the QZE, iii) single qudit. The final protocol is motivated by experimental feasibility, and we have performed a proof of concept experimental demonstration.

\end{abstract}

\maketitle

\section{Introduction}

The quantum Zeno effect (QZE) is the phenomenon where the unitary evolution of a quantum state can be suppressed e.g. by interactions with the environment or frequently performing measurements on the state \cite{DFG74, MS77}. The QZE is of interest in various fields and topics in physics, including decaying systems, trapped cold atoms, quantum computation, (nonlinear) optics and quantum foundations \cite{W97, FMR01, SM06, FP02, FJP04, YIK01, P04, HW87}. Additionally, the QZE has been implemented in a quantum protocol using entanglement to improve a communication complexity problem (CCP) beyond classical limitation \cite{HD99}.

A CCP is typically described by two parties Alice and Bob who aim to compute the value of a function $f(x,y)$, depending on bit-strings $x,y$, one given to Alice and one to Bob, despite restricted communication. The task is to maximize the probability that Bob computes $f$, given that Alice may communicate no more than $k$ bits to Bob. Naturally, this allows for various generalizations e.g. CCPs involving many parties or high-level inputs.

It is well known that quantum entanglement can give rise to correlations that do not admit a local model \cite{B64} and that such correlations can be implemented with local information processing to improve CCPs beyond classical protocols \cite{CB97, BCD01, BDHT99, BCMW10}. Close links have been established between Bell inequality violation and reduction of communication complexity \cite{BZZ02, BZPZ04, BCMW10}, providing a fundamental understanding for why quantum entanglement achieves complexity reduction.

Interestingly, it has been shown that quantum reduction of communication complexity can be achieved without shared entangled states. In \cite{TSBBZW05}, quantum strategies for two particular CCPs, relying only on sequential communication of a single quantum two-level system (qubit), were shown to lower complexity beyond classical limitations. Indeed, the superiority of these quantum strategies can no longer be explained by violations of Bell inequalities. Also in other communications tasks, where the quantum advantages previously stemmed from Bell inequality violations, has it been shown that the single qubit can be used to outperform classical protocols \cite{CTW05}. The single qubit protocols are not only conceptually interesting but also tend to be more scalable than the protocols based on entanglement, making them feasible for experimental implementations.

In this paper, we introduce a family of CCPs, generalizing the particular CCP in \cite{HD99}, and investigate classical and quantum solutions. In particular, we aim to answer two questions (1) how does entanglement and a single quantum $d$-level system (qudit) perform as resources for reduction of communication complexity?, and (2) do protocols that use the QZE perform better than protocols that do not? In our investigation, we will provide three quantum solutions to our CCPs, (i) using entanglement and the QZE, (ii) using a single qudit and the QZE, and finally (iii) using single qudit without the QZE. The final protocol is argued to be experimentally feasible and scalable, and we will provide an experimental demonstration of quantum reduction of communication complexity.

\section{Solving the Communication Complexity Problems}

The class of CCPs we consider is described as follows: imagine that a distributor supplies $M$ parties $R_1,...,R_M$ with inputs in such a way that $R_1$ and $R_M$ are given $x_1$ and $y_M$ respectively, whereas parties $R_2,...R_{M-1}$ are given two inputs each, $x_{i},y_{i}$. All inputs are elements of the set $\{0,...,dN-1\}$ where $d>1$ and $N$ is a large integer, which are publicly announced by the distributor. In addition, for $l=0,...,d-1$ define sets $S_l=\{Nl-\mu,...,Nl+\mu\}$, where addition is taken modulo $dN$, with some publicly announced positive integer $\mu$ such that $N\gg \mu $. The distributor promises the parties that each pair of inputs $x_i,y_{i+1}$ obeys $x_i-y_{i+1}\in S_l$ for some $l$. The CCP facing the $M$ parties is for $R_i$ to communicate at most one $d$-level of information to $R_{i+1}$ for $i=1,...,M-1$, in such a way that $R_M$ with high probability can announce a value $l'\in\{0,...,d-1\}$ such that $\sum_{i=1}^{M-1}x_i-y_{i+1} \in S_{l'}$. Thus, each CCP is characterized by the numbers $(N,M,d,\mu)$.

\subsection{Classical solution}

\begin{figure}
\centering
\includegraphics[width=0.9\columnwidth]{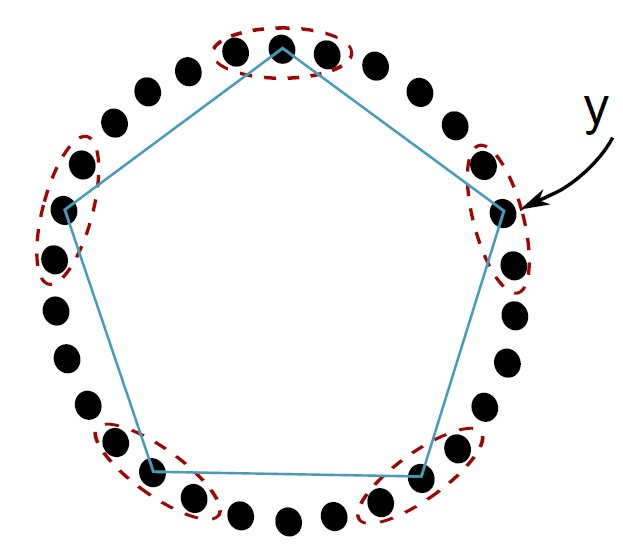}
\caption{The $dN$ numbers can be imagined as dots on a circle. For given $y$, the allowed values of $x$ are given by the $\mu$ numbers on either side of the dots that constitutes vertices of the regular convex polygon with one vertex at $y$. Here we have assumed $N=6$, $\mu=1$ and $d=5$.}
\label{figure1}
\end{figure}

Let us begin with considering classical strategies for our family of CCPs. This problem was considered for $d=2$ and $\mu=1$ in \cite{HD99} and we now solve the more general classical problem. Firstly, let us consider only two parties involved. Since the distribution of the inputs in the CCP are given on advance, we can without loss of generality restrict to deterministic classical strategies, i.e. strategies that are not randomized according to some pre-established rule. $R_1$ must then use a function $f:\{0,...,dN-1\}\rightarrow \{0,...,d-1\}$ to determine the value of the classical $d$-level that is sent to $R_2$. The $dN$ possible values of the input $x$ can be represented as dots forming a circle. Due to the distributor's promise, for a given input $y$ of $R_2$, there are only $d(2\mu+1)$ allowed possible values of $x$, all equally probable. For given $y$, the possible values of $x$ can be represented by the $\mu$ dots on either side of the particular elements constituting vertices of the regular convex polygon with $d$ sides that has one vertex on the dot representing $y$, see figure \eqref{figure1}.
$R_1$ divides the circle into $d$ equal parts, each containing $N$ subsequent dots. This division can be made arbitrarily since the distribution of the inputs is uniform. For simplicity, make divisions $G_k=\{Nk,...,N(k+1)-1\}$ for $k=0,...,d-1$. $R_1$ uses $f(x)=k$ if and only if $x\in G_k$. It is evident that whenever all the elements $y-\mu,...,y+\mu\in G_k$ for some $k$, $f(x)$ will always give $R_2$ the information necessary to with certainty find $l'$ such that $x-y\in S_{l'}$. However, whenever not all of the elements $y-\mu,...,y+\mu$ are members of $G_k$ for some $k$, then there is a probability that $R_2$ does not manage to find $l'$. The error rate depends on how many of the $2\mu+1$ numbers that are inside some $G_k$. Evidently, if we have the first $q\in\{\mu+1,...,2\mu+1\}$ numbers in $G_k$ and the final $2\mu+1-q$ numbers in $G_{k+1}$, $R_2$ will guess that $x\in G_k$ and have a probability of $\frac{q}{2\mu+1}$ to succeed. From the $dN$ possible values of $y$, only $2d\mu$ choices introduce an error. Therefore, the average probability of error is found from
\begin{equation}
P^C_{error}=\frac{1}{dN}\left(2d \sum_{i=1}^{\mu}\left(1-\frac{\mu+i}{2\mu+1}\right)\right)=\frac{\mu(\mu+1)}{N(2\mu+1)}
\end{equation}
This constitutes the error probability for two parties executing the protocol. However, in the classical strategy, this procedure is simply repeated $M-1$ times, once between each pair $R_i$ and $R_{i+1}$ of the $M$ parties, with each party $R_i$ adding to his input the received value of its previous execution of the protocol with $R_{i-1}$. The protocol succeeds whenever there is either no error in any of the executions, or if the errors cancel each other out. For large $N\gg M$, we can neglect the possibility of canceling errors and thus the error probability for $M$ parties could be estimated by
\begin{equation}\label{pclassical}
P_{error}^C(M)\approx \frac{\mu(\mu+1)(M-1)}{N(2\mu+1)}
\end{equation}
We can conclude that the probability of error is proportional to $1/N$. However, if we have $M>N$, we cannot neglect the possibility of errors canceling so \eqref{pclassical} breaks down. In analogy with the argument in \cite{HD99}, the success probability the protocol drops to the vicinity of $1/d$ which is the trivial strategy where $R_M$ takes a random guess on the answer to the CCP. To our knowledge, a proof of the optimal classical strategy is not known.

\subsection{Quantum solution with entanglement}

Let us investigate quantum strategies, and firstly let consider the protocol based on entanglement which exploits the QZE and let us called it $P_E$. This protocol is the generalization of the know result of \cite{HD99}. Let $R_1$ and $R_2$ share the entangled state $|\psi\rangle=\frac{1}{\sqrt{d}}\sum_{k=0}^{d-1}|kk\rangle$. $R_1$ will use his input $x_1$ to locally perform a unitary transformation $U(x_1)$ on his part of the shared state.
\begin{equation}\label{unitary}
U(x_1)=\sum_{k=0}^{d-1}\omega^{\frac{x_1}{N}k}|k\rangle\langle k|
\end{equation}
where $\omega=e^{\frac{2\pi i}{d}}$. Similarly, $R_2$ locally performs $U(-y_2)$. Then, $R_1$ and $R_2$ perform local measurements of their part of the state in the Fourier basis given by $|e_l\rangle=\frac{1}{\sqrt{d}}\sum_{k=0}^{d-1}\omega^{kl}|k\rangle$. The resulting probability distribution over the respective outcomes $l_1^{(R)}$ and $l_2^{(L)}$ of $R_1$ and $R_2$ is
\begin{multline}\label{prob}
P(l_1^{(R)}, l_2^{(L)})=\\
\frac{1}{d^3}\Bigg[\left(\sum_{k=0}^{d-1}\cos\left(\frac{2\pi k}{d}\left(\frac{x_1-y_2}{N}-l_1^{(R)}-l_2^{(L)}\right)\right)\right)^2\\
+\left(\sum_{k=0}^{d-1}\sin\left(\frac{2\pi k}{d}\left(\frac{x_1-y_2}{N}-l_1^{(R)}-l_2^{(L)}\right)\right)\right)^2\Bigg]
\end{multline}
However, due to the promise from the distributor we can write $x_1-y_2=aN+b$ with $a\in\{0,...,d-1\}$ and $b\in\{-\mu,...,\mu\}$. Since $\mu\ll N$, outcomes of the form $l_1^{(R)}+l_2^{(L)}=a\mod{d}$ occur with high probability, and we can choose $l_1^{(R)},l_2^{(L)}$ in $d$ different ways such that this relation is satisfied. To bound $P(l_1^{(R)}+l_2^{(L)}=a)$ from below, we neglect the contribution from the sine terms, put $b=\pm \mu$, decouple the cosine expression from the sum by putting $k=d-1$ and use the small angle approximation $\cos z\approx 1-z^2/2$, thus finding
\begin{multline}\label{boundup1}
P(l_1^{(R)}+l_2^{(L)}=a)\geq \cos^2\left(\frac{2\pi \mu(d-1)}{dN}\right)\\
\approx 1-\frac{4\pi^2 \mu^2(d-1)^2}{d^2N^2}
\end{multline}
Hence, let $R_1$ communicate his outcome $l_1^{(R)}$ to $R_2$, who computes $l_1^{(R)}+l_2^{(L)}\mod{d}$ and then with a high probability knows that $x_1-y_2\in S_{l_1^{(R)}+l_2^{(L)}}$. This accounts for the first step in the protocol. However, it is easily realized that each pair of subsequent parties $R_i$ and $R_{i+1}$ can use the above protocol with their inputs $x_i,y_{i+1}$ and $R_i$ adding his outcome $l_i^{(R)}$ to the message received from $R_{i-1}$ and communicating this to $R_{i+1}$. Thus, after the required $M-1$ executions of the protocol, $R_M$ will be able to announce the correct value $l'$ such that $\sum_{i=1}^{M-1}x_i-y_{i+1}\in S_{l'}$ if and only if there was either no error in any of the $M-1$ executions, or if the errors cancel each other out. However, for errors to cancel there must be at least two failed steps. Similar to the above classical case, we consider large $N\gg M$ and thus neglect cases with more than one error occurring allowing us to approximate the success probability as $P(l_1^{(R)}+l_2^{(L)}=a)^{M-1}$. To good accuracy, we can bound the quantum success probability from below by considering only the two first terms i.e.
\begin{equation}\label{success}
P^Q_{ent}\geq 1-\frac{4\pi^2\mu^2 (d-1)^2(M-1)}{d^2N^2}
\end{equation}
This shows that the error probability in the CCP characterized by the tuple $(N,M,d,\mu)$ is proportional to $1/N^2$, thus lowering communication complexity beyond what was achieved with the classical protocol.

\subsection{Quantum solution with single qudit}

Let us now consider quantum solutions to the CCP using only a single qudit. Our protocol $P_1$ will rely on sequential communication of a single qudit with only the final recipient performing a measurement. Thus, $P_1$ is not a manifestation of the QZE since we are not suppressing the unitary evolution with repeated measurements. Nevertheless, we will see that complexity can be reduced beyond classical limitations.

Let $R_1$ prepare the uniform superposition state $|\psi_0\rangle=\frac{1}{\sqrt{d}}\sum_{k=0}^{d-1}|k\rangle$. Now, $R_1$ acts with the local unitary $U(x_1)$ (used in protocol $P_E$) on $|\psi_0\rangle$ and then communicates the qudit to $R_2$. $R_2$ transforms the state by applying $U(x_2-y_2)$ and communicates the qudit to $R_3$. Parties $R_3,...,R_{M-1}$ act in analogy with $R_2$. When the quantum system is given to $R_M$, he performs $U(-y_M)$. The final state of the system is $|\psi_{M}\rangle=\frac{1}{\sqrt{d}}\sum_{k=0}^{d-1}\omega^{\frac{k}{N}\sum_{i=1}^{M-1}(x_i-y_{i+1})}|k\rangle$. Finally, $R_M$ performs a measurement on the qudit in the Fourier basis. The outcome of $R_M$'s measurement is labeled $l$ and is subject to a probability distribution
\begin{multline}\label{prob2}
P(l)=
\frac{1}{d^2}\Bigg[\left(\sum_{k=0}^{d-1}\cos\left(\frac{2\pi k}{d}\left(\frac{1}{N}\sum_{i=1}^{M-1}(x_i-y_{i+1})-l\right)\right)\right)^2\\
+\left(\sum_{k=0}^{d-1}\sin\left(\frac{2\pi k}{d}\left(\frac{1}{N}\sum_{i=1}^{M-1}(x_i-y_{i+1})-l\right)\right)\right)^2\Bigg]
\end{multline}
Due to the distributor's promise, we write $x_i-y_{i+1}=a_iN+b_i$ with $a_i\in\{0,...,d-1\}$ and $b_i\in\{-\mu,...,\mu\}$. Hence we can write $\frac{1}{N}\sum_{i=1}^{M-1}(x_i-y_{i+1})=\sum_{i=1}^{M-1}(a_i+\frac{1}{N}b_i)= A+\frac{B}{N}$ where we have let $\sum a_i\equiv A$ and $\sum b_i\equiv B$. Thus, we have $A\in\{0,...,(d-1)(M-1)\}$ and $B\in\{-\mu(M-1),...,\mu(M-1)\}$. However, $A$ can be reduced modulo $d$ without loss of generality. The CCP is successfully solved if $R_M$ finds $l=A\mod{d}$. Applying approximations similar to what was used to obtain \eqref{boundup1}, we find that $P^Q_{qudit}\equiv P(l=A\mod{d})$
\begin{equation}\label{success2}
P_{qudit}^Q\geq 1-\frac{4\pi^2 (d-1)^2\mu^2(M-1)^2}{d^2N^2}
\end{equation}
Indeed, the protocol $P_1$ can reduce communication complexity beyond the classical protocol. The above lower bound on success probability differs from \eqref{success} obtained using entanglement by being quadratic in $M-1$. However, in cases where $N$ is sufficiently larger than $M$, the difference between \eqref{success2} and \eqref{success} becomes negligible. However, the drop in success probability should not be interpreted as originating from the use of a single qudit instead of entanglement as a resource, but rather as consequence of not using the QZE to suppress the unitary evolution of the state. This is a manifestation of the QZE contributing to reduction of communication complexity, in addition to the quantum resource. To support this claim, we will now construct a single qudit protocol, $P_2$, for the family of CCPs that reproduces the success probability in \eqref{success} by invoking the QZE.

In $P_2$, $R_1$ prepares the same state as in $P_1$, that is: $|\psi_0\rangle=\frac{1}{\sqrt{d}}\sum_{k=0}^{d-1}|k\rangle$, performs the unitary action $U(x_1)$, and sends the qudit to $R_2$ who performs $U(-y_2)$. $R_2$ then measures the qudit in the Fourier basis using a quantum non-demolition (QND) measurement. This renders the system in one of the $d$ states forming the Fourier basis. Then, $R_2$ applies applies $U(x_2)$ to the system and communicates the state to $R_3$. Parties $R_3,...,R_{M-1}$ act in analogy with $R_2$ i.e. each will perform a unitary rotation, a QND measurement in the Fourier basis, and finally another unitary rotation after which they communicate the system to the subsequent party. When $R_M$ obtains the qudit, he performs $U(-y_M)$ and then measures (need not be a QND measurement) the system in the Fourier basis, and uses the outcome to with high probability solve the CCP. The use of the QZE plays an important role in the efficiency of protocol $P_2$. Since every pair of neighboring parties perform some unitaries on the system and then do a QND measurement, we can imagine $P_2$ as each pair of neighbors running the protocol $P_1$ throughout the line of all $M$ parties. That is, the probability of the outcome of the first QND measurement, performed by the second party in the protocol, to be associated to a successful outcome amounts to putting $M=2$ in \eqref{success2}. Since with $M=2$, \eqref{success2} becomes equivalent to \eqref{success}, the success probability of $P_2$ when run for $M$ parties becomes the equivalent to what was obtained using entanglement by invoking the same approximations.

\subsection{Comparing the protocols}
Let us now compare the strengths and weaknesses of the three quantum protocols. Indeed, since $P_1$ does not exploit the QZE, it performs worse than the other two protocols. However, both single qudit protocols but in particular $P_1$, are subject to various advantages over the protocol $P_E$ using entanglement. Firstly, $P_E$ requires the preparation and distribution of $M-1$ two-qudit entangled states whereas in $P_1$ and $P_2$ only the preparation of the uniform superposition is required. Secondly, $P_E$ requires $2(M-1)$ measurements, $P_2$ requires $M-1$ measurements, whereas $P_1$ requires only one single measurement independent of $M$. In the highly realistic case of parties having non-ideal detectors with efficiency $\eta\in[0,1]$, it is sufficient that one single measurement fails in order for the success probability to drop to the vicinity of $1/d$, which can be reproduced by guessing. The probability of all measurements succeeding in $P_E$ is $\eta^{2(M-1)}$ and in $P_2$ it is $\eta^{M-1}$, both rapidly decreasing as $M$ increases. However, using $P_1$, the probability of successful detection is constantly $\eta$. These experimental advantages make $P_1$ an experimentally feasible and scalable protocol. In table \eqref{table1} we list the properties of our protocols $P_E, P_1$ and $P_2$.
\begin{table}[t]
\centering
\begin{tabular}{|c | c | c | c |}
\hline
- & $P_E$ & $P_1$ & $P_2$ \\ [0.5 ex] % inserts table
%heading
\hline
Quantum resource & entanglement & single qudit & single qudit \\ \hline

Use of QZE & Yes & No & Yes \\\hline

Protocol efficiency & $\eta^{2(M-1)}$ & $\eta$ & $\eta^{M-1}$ \\ \hline

 Probability of failure & $ \propto M/N^2$  & $ \propto M^2/N^2$ & $\propto M/N^2$ \\ \hline
\end{tabular}
\caption{Review of the properties of the three quantum protocols $P_E$, $P_1$ and $P_2$ solving the family of CCPs.}
\label{table1}
\end{table}

\section{Experimental realization}
We will now experimentally implement our protocols $P_1$ and $P_2$. For the experimental proof of principle we have implemented the CCP protocols with $(N,M,d,\mu,)=(60,3,2,1)$ i.e. for three parties (Alice, Bob, and Charlie) using single qubit communication. In our experiment, the physical systems are defined by single photons in a polarization setup. The basis vectors $\ket{0}$, and $\ket{1}$ correspond to finding the photon in horizontal or vertical polarization respectively. Single photons are generated from a heralded single photon source through a spontaneous parametric down-conversion (SPDC) process. The idler photon is used as trigger and detected by a single photon detector $D_T$. To exactly define the spatial and spectral properties of the signal photon, the emitted photon modes are coupled into a single mode fiber (SMF) and passed through a narrowband interference filter (F). In the experimental realization, we work in the $xz$-plane of the Bloch sphere instead of the $xy$-plane used in the presented theory. Thus, we prepare the initial the photon in $\ket{H}$, the signal photon is passing through a polarizer oriented to horizontal polarization direction.

\begin{figure}
\centering
\includegraphics[width=0.9 \columnwidth]{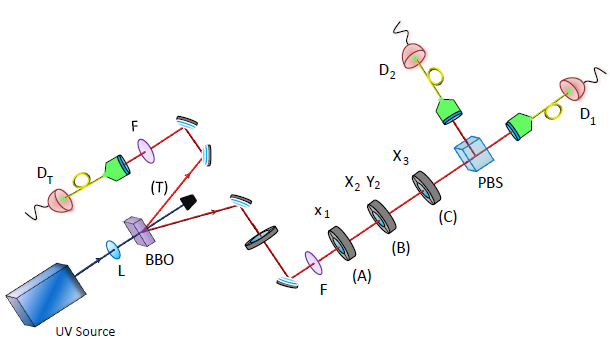}
\caption{Experimental setup for the protocol  $P_1$. Heralded single photon source consists of SPDC process, a focused (with lens L) UV light source pumping a BBO nonlinear crystal, the converted photons are emitted in two spatial modes, pass a filter (F) and coupled to single mode fiber. The idler photon is used as trigger and detected by a single photon detector $D_T$. Alice, Bob, Charlie perform their action $x_1$, $y_2$ and $x_2$, and $y_3$ by rotating half wave plates (HWP). The The polarization measurement consist of a polarization beam splitter (PBS) and two single photon detectors $D_1$ and $D_2$. These detectors are Si avalanche photodiodes (APD)}
\label{figure2}
\end{figure}
\begin{figure}
	\centering
	\includegraphics[width=0.9 \columnwidth]{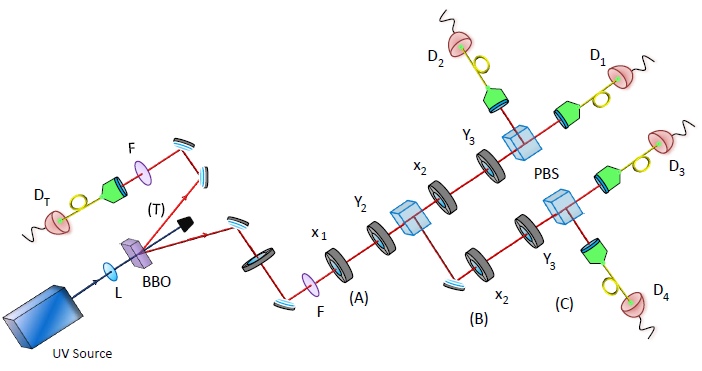}
	\caption{Experimental setup for the protocol  $P_2$. The same heralded single photon source as for $P_1$. Alice, Bob, Charlie perform their action $x_1$, $y_2$ and $x_2$, and $y_3$ by rotating half wave plates (HWP). The Bob's quantum non-demolition (QND) measurement consists  of a polarization beam splitter (PBS) where the outcome of this polarization measurement  is encoded in the path of the photon.
		 The polarization measurement consist of two polarization beam splitters (PBS) and four single photon detectors $D_1$, $D_2$, $D_3$, and $D_4$}
	\label{figure3}
\end{figure}

To execute their actions $x_1$, $y_2$ and $x_2$, and $y_3$, Alice, Bob, Charlie perform sequentially unitary transformations on the incoming qubit respectively: a rotation about the $y$-axis of the Bloch sphere with angle $\theta=\frac{\pi z_1}{N}$ where $z_1\in\{x_1,x_2+y_2,y_3\}$ for the protocol $P_1$ (see Fig.~\ref{figure2}) and with angle $\theta=\frac{\pi z_2}{N}$ where $z_2\in\{x_1,y_2,x_2,y_3\}$ for the protocol $P_2$ (see Fig.~\ref{figure2}). These transformations are achieved by rotating the polarization of the single photon with help of half wave plates (HWP). Bob's quantum non-demolition (QND) measurement, necessary for protocol $P_2$, is performed  with a polarization beam splitter (PBS) where the outcomes of this measurement (that only
addresses the polarization degrees of freedom) is encoded in the path of the photon. Finally, the measurement consists of a PBS and two single photon detectors $D_1$ and $D_2$ and two PBS and four single photon detectors $D_1$, $D_2$, $D_3$ and $D_4$ for the protocols $P_1$ and $P_2$ respectively. The $D_i$ detectors  $(i=1,2,3,4,T)$ are Si avalanche photodiodes (APD).
All coincidence counts between the signal and idler photons are registered using an three-channel coincidence logic with a time window of $1.7$~ns. The number of detected photons was approximately $9.1\times 10^4$ per second and the total time used for each experimental settings was $60$~s.
The experimental results are presented in Table~\ref{table2}. The errors come from Poissonian counting statistics and systematic errors. Due to the high photon counts, the Poissonian errors are negligible. The main sources of systematic errors are the slight intrinsic imperfections of the PBSs and HWPs.

The results are in very good agreement with the predictions of quantum mechanics. For an ideal experiment with protocol $P_1$ ($P_2$) the quantum success probability is $P_Q \geq 0.9890$ ($P_Q\geq 0.9945$) whereas the classical success probability is $P_C = 0.9778$. Our data gives the average quantum success probability $P_Q^{exp,1}\approx 0.9914\pm0.0003$ for protocol $P_1$ and $P_Q^{exp,2}\approx 0.9921\pm 0.0003$ for protocol $P_2$. Both averages are clearly above the classical bound and our experiment demonstrates the advantage of  protocol $P_2$ over $P_1$.

\begin{table}[t]
\begin{tabular}{ |c|cccccc|} \hline\cline{1-7}

c$ No $&$x_{1}$ & $y_{2}$ & $x_{2}$ & $y_{3}$ & $P_{success}^{exp,1}$ & $P_{success}^{exp,2}$  \\ 
\hline\cline{1-6}
% THE ERRORS ARE STILL THE OLD ONES
\: 1 & 70 &71  & 55  & 56  & 0.9944$\pm 0.0004 $ & 0.9927$\pm0.0021$ \\
\: 2 & 58 &59  & 38  & 37  & 0.9800$\pm 0.0035$  & 0.9951$\pm0.0021$  \\         
\: 3 &67 & 7  & 88  & 29  &  0.9864$\pm 0.0019$ & 0.9894$\pm 0.0015$ \\
\: 4 &40 &101 & 15  & 76  &  0.9904$\pm 0.0004 $ & 0.9948$\pm 0.0021$\\
\: 5 &70 & 10 & 40  & 101 &  0.9967$\pm 0.0019 $  & 0.9918 $\pm 0.0015$\\
\: 6 &36 & 36 & 117 & 56  &  0.9812$\pm 0.0019 $  & 0.9896 $\pm 0.0015$\\
\: 7 &44 &103 & 117 & 57  &  0.9965$\pm 0.0018 $  & 0.9900 $\pm 0.0015$\\
\: 8 &80 &19  & 108 & 108 &  0.9983$\pm 0.0019 $  & 0.9931 $\pm 0.0015$\\
\: 9 &36 &36  & 38  & 37  &  0.9797$\pm 0.0019 $ & 0.9951 $\pm 0.0015$\\
\: 10 &117 &57  &  72 & 13 & 0.9814$\pm 0.0018 $ & 0.9915 $\pm 0.0015$\\
\: 11 &0  & 0  &  63 &  4  & 0.9884$\pm 0.0018 $ & 0.9927 $\pm 0.0015$\\
\: 12 &17 &18  &  67 &  67 & 0.9852$\pm 0.0018 $ & 0.9980 $\pm 0.0015$\\
\: 13 &99 &40  &  19 &  80 & 0.9963$\pm 0.0004 $ & 0.9934 $\pm 0.0021$\\
\: 14 &79 &18  &  80 &  79 & 0.9900$\pm 0.0004 $ & 0.9869 $\pm 0.0021$\\
\: 15 &48 &108 &  69 &  10 & 0.9978$\pm 0.0019 $ & 0.9939 $\pm 0.0016$\\
\: 16 &104 &105 &  8 &   7 & 0.9836$\pm 0.0036 $ & 0.9897 $\pm 0.0021$\\
\: 17 &25 &25 &  20 &  80  & 0.9902$\pm 0.0004 $  & 0.9956 $\pm 0.0003$\\
\: 18 &33 &94 &  59 &  118 & 0.9878$\pm 0.0036 $  & 0.9960 $\pm 0.0022$\\
\: 19 &87 &28 &  40 &  100 & 0.9849$\pm 0.0018 $  & 0.9957 $\pm 0.0016$\\
\: 20 &63 & 3 &  98 &   38 & 0.9926$\pm 0.0004 $  & 0.9914 $\pm 0.0003$\\
\: 21 & 119  &  58 & 115 & 54 & 0.9973$\pm 0.0004 $ & 0.9900 $\pm 0.0021$	\\
\: 22 & 110 &50  & 101 &  41 & 0.9971$\pm 0.0004 $ & 0.9915 $\pm 0.0004$\\   
\: 23 &  64  &   3 &  58 & 59 & 0.9991 $\pm 0.0038 $ & 0.9932 $\pm 0.0020$\\
\: 24 &  82  &  81 &  94 & 33 & 0.9909 $\pm 0.0004 $ & 0.9840 $\pm 0.0022$\\
\: 25 &  60  & 0 & 0 & 119 & 0.9843 $\pm 0.0018 $ & 0.9920 $\pm 0.0015$\\
\: 26 &  60  & 60 &  15 & 16 & 0.9989$\pm 0.0019 $ & 0.9893 $\pm 0.0015$\\
\: 27 &  108 & 47 & 119  & 0 & 0.9948$\pm 0.0038 $ & 0.9937 $\pm 0.0022$\\
\: 28 &  94  &  33 &  14 & 13 & 0.9984$\pm 0.0004 $ & 0.9939 $\pm 0.0021$\\
\: 29 & 114  & 55  &  7  &  8  & 0.9974$\pm 0.0004 $& 0.9915 $\pm 0.0021$\\
\: 30 &  109  & 49  &  9  &  8  & 0.9879$\pm 0.0020 $ & 0.9912 $\pm 0.0016$\\
\: 31 & 103  & 103 &  90 & 29 & 0.9962$\pm 0.0019 $ & 0.9915 $\pm 0.0015$\\
\: 32 &  74  & 73 &  24 & 85 & 0.9980$\pm 0.0037 $ & 0.9902 $\pm 0.0021$\\
\: 33 &   2  & 63 & 28 & 28 & 0.9877$\pm 0.0018 $ & 0.9852 $\pm 0.0016$\\
\: 34 & 109  & 49 & 69 & 8 & 0.9852$\pm 0.0019 $ & 0.9925 $\pm 0.0015$\\
\: 35 &  7  & 7 & 44 & 44 & 0.9874$\pm 0.0004 $ & 0.9923 $\pm 0.0003$\\
\: 36 &  110  & 50 & 90 & 30 & 0.9960$\pm 0.0004 $& 0.9901 $\pm 0.0004$\\
\: 37 &   56  & 56 & 5 & 64 & 0.9849$\pm 0.0019 $ & 0.9914 $\pm 0.0015$\\
\: 38 &    9  &  9 & 11 & 71 & 0.9871$\pm 0.0004 $ & 0.9965 $\pm 0.0003$\\
\: 39 &  48   & 49 &  6 & 67 & 0.9946$\pm 0.0004 $ & 0.9953 $\pm 0.0020$\\
\: 40 &  2  & 2 & 98 & 97 & 0.9891$\pm 0.0019 $ & 0.9907 $\pm 0.0015$\\
\hline\cline{1-7}
\end{tabular}
  \caption{Experimental results for the success probability with inputs $x_{1}$, $y_{2}$ and $x_{2}$, and $y_{3}$ for Alice, Bob, and Charlie respectively. By $P_{success}^{exp,1}$ ($P_{success}^{exp,2}$) we denote the measured results for protocol $P_1$ ($P_2$).} 
\label{table2}
\end{table}

\section{conclusions}
In this paper, we have investigated classical and quantum solutions for a family of CCPs. We provided a classical solution and proposed three different quantum protocols improving the CCPs beyond the classical performance. Two of the quantum protocols use QZE, one relying on entanglement as a resource while the other relying on single qudit communication, and we showed that the performance of both protocols is equal. We also proposed a protocol with single qudit communication without the QZE, and the performance was shown to be lower than the other two quantum protocols. We gave a proof of concept experimental demonstration of reduction of communication complexity beyond the classical bound for both the single qudit protocol using the QZE and the protocol that does not use the QZE. Our experimental findings demonstrated the advantages of using the protocol based on the QZE. Our results suggest that the use of the QZE together with quantum resources could enhance information processing in certain tasks in comparison to quantum protocols not using the QZE. 

This project was supported by the Swedish Research Council, ADOPT,  and QOLAPS project of ERC.


\begin{thebibliography}{99}


\bibitem{MS77} B. Misra and E. C. G. Sudarshan, \textit{The Zeno's Paradox in Quantum Theory}, J. Math. Phys. \textbf{18}, 756 (1977).

\bibitem{DFG74} A. Degasperis, L. Fonda and G. C. Ghirardi, \textit{Does the lifetime of an unstable system depend on the measuring apparatus?}, II Nuovo Cimento A \textbf{21},3 pp. 471-484 (1974).


\bibitem{W97}
S. R. Wilkingson, et al., \textit{Experimental evidence for non-exponential decay in quantum tunnelling}, Nature \textbf{387}, 575-577 (1997).

\bibitem{FMR01}
M. C. Fisher, B. Gutiérrez-Medina and M. G. Raizen,\textit{Observation of the quantum Zeno and anti-Zeno effects in an unstable system},
Phys. Rev. Lett. \textbf{87}, 040402 (2001).

\bibitem{SM06}
E. W. Streed, et al., \textit{Continuous and pulsed quantum Zeno effect}
Phys. Rev. Lett. \textbf{97}, 260402 (2006).

\bibitem{FP02}
P. Facchi and S. Pascazio, \textit{Quantum Zeno subspaces}, Phys. Rev. Lett. \textbf{89}, 080401 (2002).

\bibitem{FJP04}
J. D. Franson, B. C. Jacobs and T. B. Pittman, \textit{Quantum computing using single photons and the Zeno effect}, Phys. Rev. A \textbf{70}, 062302 (2004).

\bibitem{YIK01}
K. Yamane, M. Ito and M. Kitano, \textit{Quantum Zeno effect in optical fibers},
Optics Communications \textbf{192}, 3-6 (2001).

\bibitem{P04}
J. Perina, \textit{Quantum Zeno effect in cascaded parametric down-conversion with losses},
Optics Communications \textbf{325}, 1 (2004).

\bibitem{HW87}
D. Home and M. A. B Whitaker, \textit{The many-worlds and relative states interpretations of quantum mechanics, and the quantum Zeno paradox},
J. Phys A: Math. Gen. \textbf{20} 3339 (1987).

\bibitem{HD99} L. Hardy and W. van Dam, \textit{Quantum communication using a nonlocal Zeno effect
}, Phys. Rev. A \textbf{59}, 2635 (1999).


\bibitem{B64} J. S. Bell, \textit{On the Einstein Podolsky Rosen paradox}, Physics, \textbf{1}, 195 (1964).
	
\bibitem{CB97} R. Cleve and H. Buhrman, \textit{Substituting quantum entanglement for communication}, Phys. Rev. A \textbf{56}, 1201 (1997).
	

\bibitem{BCMW10} H. Buhrman, R. Cleve, S. Massar and R. de Wolf, \textit{Nonlocality and communication complexity}, Rev. Mod. Phys.  \textbf{82}, 665 (2010).


\bibitem{BCD01} H. Buhrman, R. Cleve and W. van Dam, \textit{Quantum entanglement and communication complexity}, SIAM J.Comput \textbf{30} 1829-1841 (2001).

\bibitem{BDHT99} H. Buhrman, W. van Dam, P. Hoyer and A. Tapp, \textit{Multiparty quantum communication complexity}, Phys. Rev. A \textbf{60}, 2737 (1999).

\bibitem{BZZ02} C. Brukner, M. Zukowski and A. Zeilinger, \textit{Quantum communication complexity protocol with two entangled qutrits}, Phys. Rev. Lett. \textbf{89}, 197901 (2002).

\bibitem{BZPZ04} C. Brukner. M. Zukowski. J-W. Pan and A. Zeilinger, \textit{Bell’s inequalities and quantum communication complexity
}, Phys. Rev. Lett. \textbf{92}, 127901 (2004).


\bibitem{TSBBZW05} P. Trojek, C. Schmid, M. Bourennane, C. Brukner, M. Zukowski and H. Weinfurter, \textit{Experimental quantum communication complexity
}, Phys. Rev. A \textbf{72}, 050305(R) (2005).

\bibitem{CTW05} C. Schmid, P. Trojek, H. Weinfurter, M. Bourennane, M. Zukowski and C. Kurtsiefer, \textit{Experimental single qubit quantum secret sharing}, Phys. Rev. Lett. \textbf{95}, 230505 (2005).






\end{thebibliography}
\end{document}